\documentclass{nature}
\usepackage{graphicx} 
\usepackage{cite}
\usepackage{epsfig} 
\usepackage{amsmath}
\usepackage{xcolor}
\usepackage{soul}
\usepackage{ulem}
\usepackage{caption}
\usepackage{xr}

\title{Ultra-low-loss optical interconnect enabled by topological unidirectional guided resonance}

\author{Haoran Wang$^{1}$, Yi Zuo$^{1}$,  Xuefan Yin$^{4}$, Zihao Chen$^{1}$, Zixuan Zhang$^{1}$, Feifan Wang$^{1}$, Yuefeng Hu$^{2,3}$, Xiaoyu Zhang$^{1}$ \& Chao Peng$^{1,2*}$}
\begin{document}

\maketitle

\begin{affiliations}
\item State Key Laboratory of Advanced Optical Communication Systems and Networks, Department of Electronics \& Frontiers Science Center for Nano-optoelectronics, Peking University, Beijing 100871, China
\item Peng Cheng Laboratory, Shenzhen 518055, China
\item Peking University Shenzhen Graduate School, Shenzhen 518055, China
\item Department of Electronic Science and Engineering, Kyoto University,
Kyoto-Daigaku-Katsura, Nishikyo-ku, Kyoto 615-8510, Japan
\end{affiliations}
Corresponding Author: Chao Peng (email: pengchao@pku.edu.cn)

\begin{abstract}
Grating couplers that interconnect photonic chips to off-chip components are of essential importance for various optoelectronics applications. Despite numerous efforts in past decades, existing grating couplers still suffer from poor energy efficiency and thus hinder photonic integration toward a larger scale. Here, we theoretically propose and experimentally demonstrate a method to achieve ultra-low-loss grating coupler by employing topological unidirectional guided resonances (UGRs). Leveraging the unidirectional emitting nature of UGRs, the useless downward radiation is greatly suppressed with no mirror placed on the bottom. By engineering the dispersion and apodizing the geometry of grating, we realize a grating coupler on 340 nm silicon-on-insulator platform with a record-low-loss of 0.34 dB and  bandwidth exceeding 30 nm at the telecom wavelength of 1550 nm. We further show a pair of grating couplers works as optic via that interconnects two stacked photonic chips with a loss of only 0.94 dB. Our work sheds light on the feasibility of energy-efficient optical interconnect for silicon photonics, and paving the way to large-scale photonic integration for applications from optical communication to photonic computing.

\end{abstract}

\section{Introduction}
Optical I/Os are indispensable building blocks for silicon photonics since they are responsible for coupling the light into and out of the chip\cite{sun2015single,marchetti2019coupling,cheng2020grating,nambiar2018grating,carroll2016photonic,xiang2021perspective,urino2014high,tasker2021silicon,yang2022multi,chen2017flexible,zhang2022large,huang2023leaky,atabaki2018integrating,zhang2023photonic,spencer2018optical,miller2020large,li2020silicon}. To tackle higher coupling efficiency, a variety of techniques such as micro-optics\cite{scarcella2017pluggable,dietrich2018situ}, edge coupling\cite{hatori2014hybrid,cheben2015broadband,papes2016fiber}, grating coupling\cite{notaros2016ultra,wade201575,benedikovic2015high,benedikovic2017shaped,chen2017experimental,chen2017dual,tian2022high,wan2016grating,wan2018grating,wan2017waveguide,sodagar2014high,xu2019sin,zaoui2014bridging,ding2014fully,hong2019high,nambiar2019high,hoffmann2018backscattering,marchetti2017high,schmid2009optimized,ding2013ultrahigh,tseng2015high,haglund2019high,lu2016flip,rogers2021universal},and photonic wire bonding\cite{lindenmann2012photonic,lindenmann2014connecting} are proposed and investigated. Among these techniques, the grating coupling is a particularly promising method owing to its small footprint, great flexibility in arrangement, and potential in mass production\cite{carroll2016photonic,nambiar2018grating,cheng2020grating,marchetti2019coupling}. However, despite extensive efforts, existing grating couplers still suffer from poor energy efficiency that would significantly degrade the power budget of the optical links, and thus hinders its application in large-scale photonic integration for next-generation optoelectronic system-on-chip (SoC) that needs to drive tens of Tbps of bandwidth directly into or out of high-performance ICs through a massive of optical interconnects on a single chip. Because conventional gratings radiate both upwards and downwards, nearly half of the light wastes at the lower substrate, and consequently, causes an intrinsic loss of $\sim$3 dB unless a mirror is placed at the bottom\cite{zaoui2014bridging,ding2014fully,hong2019high,nambiar2019high}. Recently, unidirectional radiation without mirrors has been discovered from the view of topology. It is found that a class of unidirectional guided resonances (UGRs)\cite{yin2020observation,yin2023topological,zhang2022all,zhang2021observation} can be realized by merging a pair of half-integer topological charges upon one side of the grating while leaving them apart on the other side\cite{yin2020observation,yin2023topological}. The UGRs are proved to be ubiquitous in periodic photonic structures since the half-charges can be either created from splitting an integer charge carried by a bound state in the continuum (BIC)\cite{hsu2016bound,hsu2013observation,zhen2014topological,doeleman2018experimental,zhang2018observation,wang2022fundamentals,price2022roadmap,yin2020manipulating,peng2020trapping}, or alternatively, spawned from a  void point owing to the inter-band coupling effect\cite{yin2020observation,yin2023topological}. UGRs shed light on the new possibilities of unidirectional emission, while how to utilize them to construct a complete, practical, and fabrication-friendly grating coupler to support ultra-low-loss optical interconnect remains an important but elusive problem.

Here, we theoretically propose and experimentally demonstrate a method to realize ultra-low-loss grating couplers by utilizing the unidirectional emission nature of topological UGRs\cite{yin2020observation,yin2023topological}. We adopt an L-shaped structure that is simple and planar-process-compatible and show the integer topological charge carried by a symmetry-protected BIC splits into a pair of half-charges. By continuously tuning the grating geometry, the half-charges evolve in the momentum space and restore integer charge at the lower side of the grating to form a UGR. The design is optimized by engineering the dispersion and apodizing the geometry to best reduce back-scattering and promote the model overlap with the fiber\cite{zhao2020design}. We fabricate the grating coupler samples on a 340 nm thick SOI wafer and obtain a record-low insertion loss of 0.34 dB and 0.94 dB for the schemes of chip-to-fiber and stacked-chip interconnects at telecom wavelength of 1550 nm, with their 1 dB bandwidth exceeding 30 nm and 20 nm, respectively. Our findings pave the path to ultra-high energy-efficient optical interconnects of silicon photonics and reveal the potential of photonic integration toward ultra-density and 3D-stacking applications.

\section{Design and principles}
\setcounter{equation}{0}
\renewcommand{\theequation}{\arabic{equation}}
The goal of this work is to develop an ultra-low-loss grating coupler with a sufficiently broad bandwidth by utilizing a resonance with unidirectional radiation, namely a UGR. To make the device practically useful, the design has to be applied on a standard SOI wafer and the geometry should be compatible with the planar CMOS process. To fulfill these requirements, we start with a schematic design illustrated in Fig.~1a, in which a ``L-shaped" grating is patterned on a 340 nm thick SOI wafer with a periodicity of $a=528$ nm and then buried by a standard silicon dioxide cladding layer for protection. The unit-cell design of the grating (inset, Fig.~1a) has two different widths and depths with vertical sidewalls, denoted as $w_1, h_1$ and $w_2, h_2$, respectively. Such a structure can be fabricated by simple overlay lithography and dry etch steps, to avoid sophisticated tiled etching\cite{yin2020observation} or dual-layer deposition processes\cite{notaros2016ultra,wade201575}. As reported, the key to generating the UGRs is to create a pair of half-integer topological charges ($q=\pm 1/2$) carried by circularly polarized (CP) states, which can be accomplished by splitting an integer-charge $q=\pm 1$ of tunable BIC\cite{yin2020observation}, or zero-charge $q=0$ of a void point through inter-band coupling\cite{yin2023topological}. However, such mechanisms usually require a relatively thick wafer and thus are not easy to be compatible with standard SOI thickness\cite{yin2020observation,yin2023topological,zhang2022all,zhang2021observation}.  

Alternatively, we turn to work on a symmetry-protected BIC which robustly resides at the Brillouin zone (BZ) center for any slab thickness. As shown in Fig.~1c, the grating operates at the lowest, anti-symmetric transverse electric (TE) band that we denote it as TE-A band. We treat the L-shaped hole as a combination of two rectangular blocks B1 and B2 with different widths and depths. When the small block B1 is absent, the grating restores $C_2$ symmetry and raises a symmetry-protected BIC at the $\Gamma$ point\cite{zhen2014topological}, carrying a topological charge of $q=-1$ and exhibiting as a diverged quality factor ($Q$) (black line, Fig.~1c). The presence of the small block B1 breaks the $C_2$ symmetry and splits the integer charge into a pair of half-charges $q=-1/2$. Accordingly, the $Q$s of resonances in the vicinity of the $\Gamma$ point degrade to $\sim134$ (red line, Fig.~2c) which is sufficiently low to support broadband operation. A UGR is found at $k_x=-0.1192$  with an asymmetry ratio (defined as the ratio between upward and downward radiation intensity) reaching $\eta\sim$ 65 dB. As confirmed by the field pattern (Fig.~1b), the UGRs unidirectionally radiate toward the upper direction while nearly no energy leaks into the lower substrate. 

The trajectories of topological charge upon the downward radiation are shown in Fig.~1d: red for right-handed circularly polarized (RCP) and blue for left-handed circularly polarized (LCP) which are opposite in helicity. By gradually increasing the depth of the small block $h_1$ from $0$ nm to $100$ nm, a pair of half-charges $q=-1/2$ split from the BIC, and evolve in a $y$-mirror symmetric manner in the momentum space. The RCP and LCP trajectories meet on the $k_x$ axis at $k_x =-0.1192$ at which $h_1=100$ nm. At this point, any downward radiation needs to be both LCP and RCP at the same time, which can never be satisfied. As a result, the grating resonance cannot have any downward radiation, even without a mirror on the bottom.

Next, we design the complete structure of the grating coupler to incorporate the UGR with waveguide and fiber modes. Fig.~2a shows the top view of the unidirectional grating coupler, which consists of a tapered-waveguide region, an apodization region, and a uniform region. The dispersion of the UGR and waveguide modes are plotted in Fig.~2b, showing that the group velocities  $v_g=d\omega/dk$ of the waveguide modes are almost constant when the waveguide width shrinks from $20~\mu$m to $2~\mu$m. By fine-tuning the unit-cell geometry of the grating, we also make the group velocity of the uniform region (red curve, Fig.~2b) almost identical to the tapered waveguides (black and gray lines, Fig.~2b). Therefore, the group-velocity-matching has been fulfilled that allows the energy to efficiently transit between the grating and waveguide modes.

To miniaturize the back-scattering between the grating and waveguide, we perform an adiabatic transformation of grating geometry to handle the momentum mismatch between the UGR and waveguide mode.  A detailed side view of the apodization region is illustrated in Fig.~2c, in which the grating periodicity $a$ and the width of blocks B1 and B2 ($w_1$, $w_2$) are continuously adjusted, marked as gradient colors. We pick up three grating geometries at different positions from A to C, and calculate their bulk dispersion, respectively (Fig.~2b). As expected, the apodization smoothly transits the UGRs to the targeted waveguide mode that fulfills the requirement of momentum-matching.

Further, we design the grating coupler to best match its radiation distribution with a standard single-mode fiber (Corning SMF-28e+) with core and cladding diameters of 8.2 $\mu$m and 125 $\mu$m. As light propagates down the fiber, the beam maintains a nearly Gaussian cross-sectional profile with a mode field diameter (MFD) of 10.4 $\mu$m (Fig.~2d).  To promote the overlap coefficient with the fiber mode, the spatial and angular distributions of the radiation field need to be tuned simultaneously. By carefully arranging the apodization region, we obtain the optimized design of the unidirectional grating coupler. Numerical simulation (Fig.~2d, Lumerical FDTD) confirms that the waveguide mode propagates through grating with minor back-scatterings, and the light only radiates upwards in an oblique direction of $\theta=13.72^\circ$. The detailed spatial and angular distributions of the radiation field are presented in Fig.~2e, showing nearly Gaussian shapes that match well with the targeted standard single-mode fiber. 

To show the bandwidth and angular tolerance performances of the unidirectional grating coupler, we calculate the map of fiber-to-coupler coupling efficiency (CE) over the wavelength and alignment angle  (Fig.~2f). The red and blue contours represent the CE thresholds of 3 dB and 1 dB, showing that the unidirectional grating coupler works in a spectrum width of 57.6 nm and 28.9 nm with angular tolerance of $6^\circ$ and $3^\circ$, respectively. The calculation confirms that the designed grating coupler has excellent robustness owing to the topological nature of the UGR, and it is capable of broadband operation and easy assembly.

\section{Sample fabrication and experimental characterization}
To verify the design and principles, we fabricate the unidirectional grating coupler samples on a standard 340-nm-thick SOI wafer by using overlaid electron-beam lithography and inductively coupled plasma etching (See Methods for more details).  Our design doesn't require sophisticated silicon deposition\cite{notaros2016ultra,wade201575} or tilted etching steps\cite{yin2020observation}, and thus significantly simplifies the fabrication process. A top-view of the grating coupler sample is observed by using an optical microscope, showing a total grating footprint of $20 \times 20 ~\mu$m (Fig.~3a).  The scanning electron microscope images of the apodization and uniform regions are presented in Fig.~3c,d, and the detailed side and top views of the ``L-shaped" grating patterns are shown in Fig.~3b,e, which gives $a=528$ nm, $w_1=158$ nm, $w_2=200$ nm, $h_1=100$ nm, $h_2=262$ nm in the fabricated samples. The underlying SiO$_2$ layer is maintained, while for the simplicity of testing, the grating coupler is immersed into a refractive index liquid of $n=1.46$ to simulate the deposited upper cladding of SiO$_2$ in the foundry process. 

To evaluate the performance of the unidirectional grating coupler, we construct a fiber-to-detector optical interconnect in which the light transmits through two identical grating couplers and the waveguide connected them, as schematically shown in Fig.~4a.  A tunable telecommunication laser with light in the C + L band is first sent through a polarization controller to match with the grating and then transmits into the fiber. We use a free-space photodiode with an active area of  $10 \times 10$ mm to receive the light (See Methods for more details.)  By sweeping the wavelength from 1510 to 1590 nm, we measure the insertion losses of the fiber-to-detector link as illustrated in Fig.~4b, showing a minimal loss of 1.24 dB at 1550 nm. 

Since the fiber and photodiode have different receiving areas and apertures, their overlap factors with respect to the coupler are also different. We calculate the overlap ratio between the coupler-to-detector and coupler-to-fiber interfaces from the devices' parameters (inset, Fig.~4b). In this test, the chip contains a waveguide in the length of 7 mm to connect the two unidirectional grating couplers, and we calibrate the waveguide losses from some reference samples (see Supplementary Section for details). As a result, we decompose the total link loss to the coupler-to-detector CE and coupler-to-fiber CE, respectively (Fig.~4b). The measured results show good agreement with the design. 
The peak CE at the coupler-to-fiber interface reaches a record-high value of 0.34 dB at 1550 nm, namely 92.47\% of light energy successfully couples into the fiber. Accordingly, the 1 dB bandwidth of the unidirectional grating coupler exceeds 30 nm, guaranteeing its capability of broadband operation.

Owing to the protection of topology, the unidirectional grating coupler is expected to maintain fairly good performance under geometry deviation\cite{yin2020observation,yin2023topological}. To evaluate and verify the fabrication tolerance, we fabricate a group of samples with the rectangular blocks B1 and B2 slightly shifting in the transverse direction which is a common type of imperfection in the overlaid lithography process (inset, Fig.~4d). Specifically, we sweep the shift from -50 nm to 50 nm in a step of 10 nm and obtain the coupler-to-fiber CE by using the approach elaborated above. As shown in Fig.~4d, the peak CEs keep higher than 1 dB across all samples and can remain better than 0.6 dB if the shift error is below $\pm$ 30 nm. We also plot the map of CE on the wavelength and shift in Fig.~4c, confirming that the ultra-low and broadband feature of the unidirectional coupler is robust under fabrication deviation. 

To reveal the robustness and repeatability of the unidirectional grating coupler again random disorders, we further fabricate 36 identical devices by applying the ideal design to the same fabrication process, in which the standard deviations of the pattern locations and widths are estimated to be about 5 nm. The measured CEs statistically show an average value of -0.426 dB with a variance of 0.005 dB (Fig.~4e). The averaged 1dB bandwidth is 32.83 nm across the samples, with a variance of 0.55 nm. 

The unidirectional emission nature of the proposed grating coupler sheds light on the possibility of 3D stacked interlaying of photonic chips in which a pair of couplers work as optical via, allowing the mediation of light energy flow across the upper and lower chips\cite{wan2016grating,wan2018grating,wan2017waveguide,sodagar2014high,xu2019sin}. To verify this concept, we perform numerical simulation (Lumerical, FDTD) in which the waveguide mode transmits into and radiates through the unidirectional grating coupler on the lower chip, received by another flip-aligned coupler, and then propagates in the upper chip (Fig.~5a). Thanks to the ultra-low-loss nature of the unidirectional grating coupler, the insertion loss of such a interlay optical interconnect can be below 1 dB to fulfill the power budget for high-speed data transmission between two 3D stacked chips.

We further experimentally demonstrate the interlay optical interconnect by using the unidirectional grating couplers. The measurement system is illustrated in Fig.~5b, in which one photonic chip is vertically flipped and overlapped on the other chip to form an optical link from the fiber to the detector, which contains 4 unidirectional grating couplers and waveguides in a total length of 10 mm (see Supplementary Section for details). We directly measure the insertion loss of the entire interlay link as shown in Fig.~5c, which gives a minimal loss of -2.68 dB at 1550 nm.  After ruling out the fiber-to-coupler, fiber-to-detector, and waveguide losses that are measured previously (see Supplementary Section for details), we obtain the coupler-to-coupler CE from the total link loss that attributes to the coupler pair only. As shown in Fig~5d, the peak CE reaches 0.94 dB at 1550 nm, indicating that 80.53\% of light energy flows across the optical via with a 1 dB bandwidth of 21.4 nm. 

\section{Discussions}
Above theory and experiments demonstrate a new method of realizing ultra-low-loss grating coupler from the perspective of topological charge manipulation. We achieve a record-low peak CE of $0.34$ dB in a standard 340 nm SOI wafer with grating shapes that are compatible with the foundry process. Our finding provides a vivid example of improving the photonic devices from the view of topology and reveals the feasibility of ultra-high energy-efficient optical interconnects of silicon photonic for dense integration and 3D stacked integration. Given that topological UGRs 
ubiquitously exist in many symmetry-broken photonic structures, the proposed technique can be applied to other material systems (such as lithium niobate on insulator platform) to improve chip-to-fiber coupling efficiencies, in which the vertical sidewalls are usually hard to achieve.

\section{Conclusion}
To summarize, we present a new method of ultra-low-loss interconnect by using a unidirectional radiating grating coupler. By splitting and then restoring an integer topological charge and apodizing the grating geometry to adiabatically match the waveguide with the fiber mode, we achieve a record-low fiber-to-chip coupling efficiency of 0.34 dB with its bandwidth exceeding 30 nm. The unidirectional grating coupler also enables an energy-efficient interconnect between two stack photonic chips, with an interlay insertion loss of only 0.94 dB.  Our work highlights the great potential of silicon photonics for the next-generation ultra-dense data transmission, and paves the way to large-volume on-chip photonic signal processing and computing.

\section*{Methods}
\noindent \textbf{Numerical simulation.} 
The COMSOL results are all calculated in the ``Radio Frequency" module in the frequency domain. The band structure and quality factor are based on the eigenvalue solver. The asymmetry ratio is solved by the two surface probes which are above and below the structure. Periodic (Floquet) boundary conditions are set along the $x$ and $y$ direction. The perfectly matching layers are set along the $z$ direction. The waveguide dispersion is based on the Lumerical mode solver. For Lumerical FDTD, the mode source is set at the left side of the grating coupler to excite the grating coupler and the perfectly matching layers are set to surround the whole region. The electrical profile is acquired through a $x$-$z$ two-dimensional monitor. The spatial distribution is extracted from the one-dimensional monitor placed 2 $\mu$m above the structure. The angular distribution is extracted from the far-field projection of the monitor. To calculate the coupling efficiency, the monitor is set at the left side of the grating coupler to collect the power from the Gaussian incidence with 10.4 $\mu$m mode field diameter.

 \vspace{3pt}
\noindent \textbf{Sample fabrication.} 
The sample is fabricated on a silicon-on-insulator wafer with a 340 nm silicon layer thickness. The grating coupler pattern of block B2 is defined by electron-beam lithography (EBL). The sample is firstly spin-coated with a 340 nm thick layer of ZEP520A photo-resist followed by being exposed to EBL (Elionix ELS-F125G8) at the current of 1 nA and 500 $\mu$m field size. After the exposure, the sample is etched with ICP (Oxford) by a mixture of {C}$_4${F}$_8$, {S}{F}$_6$ and Ar. The resist is removed with DMAC solution. To make the L-shaped grating pattern, a second round of exposure and etching is performed to fabricate block B1, with the cross-shaped alignment marks defined by the first round of exposure and etching. Next, the waveguide and tapered-waveguided regions are fabricated by overlaying another round of exposure and etching described above to etch through the silicon layer. The ICP etch times are carefully controlled for different shapes and depths.
 
 \vspace{3pt}
\noindent \textbf{Measurement system.} 
A Santec TSL-550 tunable laser generates the source from 1510 nm to 1590 nm. A Thorlabs FDG10X10 photodiode with a 10 mm$\times$10 mm large receiver area is used to fully collect the radiation energy. To measure the interlay coupling efficiency between two chips, two samples with grating arrays are stacked. A Thorlabs PT3-Z8 three-dimensional motorized translation stage is used to sweep the upper grating to the optimal position. The sweep region is reduced to the region given by the spacing of the grating array. During the measurement, the chips are immersed in a refractive index liquid of $n=1.46$ (Cargille, Series A) to simulate the SiO$_2$ cladding in the foundry process.

\section*{Conflict of interest}

The authors declare that they have no conflict of interest.


\section*{Author contributions}
C. P. conceived the idea. H. W., X.Y., and C. P. performed the theoretical study. H. W., Z. C., Z. Z, F. W.and Y. H. performed the analytical calculations and numerical simulations. H. W., Y. Z., Z. Z., F. W., Y. H. and X. Z conducted the experiments and analyzed the data. C. P. and H. W. wrote the manuscript, with input from all authors. C. P. supervised the research. All authors contributed to the discussions of the results.

\section*{References}
\bibliography{Reference}{}
\bibliographystyle{naturemag} 

\clearpage
\begin{figure}
\centering
\includegraphics[width=12 cm]{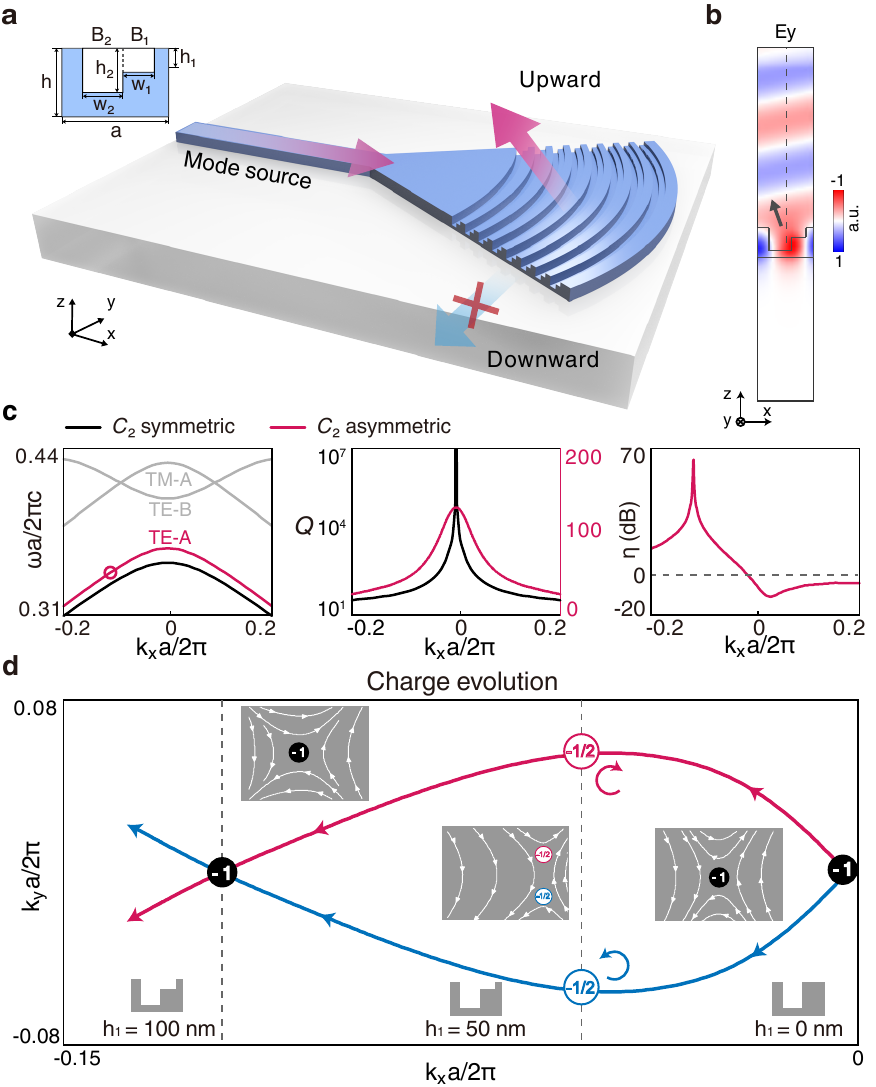}
\end{figure}
\begin{figure}
\caption{\textbf{Unidirectional guides resonance from topological charge evolution}.
\label{mfig1}
(a) 
Schematic of L-shaped grating coupler. Light travels through and radiates upwards from the grating with downward radiation being forbidden.  
(b)
The electrical field ($E_y$) of the UGR which radiates upward only.
(c)
The left panel: Band structures with the UGR marked by a circle, for $C_2$ symmetric (red) and asymmetric (black) structures. The middle panel: The quality factors of the TE bands for the symmetric and asymmetric unit cell. A symmetry-protected BIC is located at the $\Gamma$ point. By breaking the symmetry, the quality factor reduces from infinite to a friendly value of 134 for broadband operation. The right panel: The asymmetry ratio reaches 65.8 dB at $k_x$=-0.1192 which indicates a UGR.
(d)
Trajectories traced by the two half charges (red and blue) for downward radiation in momentum space by varying the height of block B1, showing that the integer charge $q=-1$ splits and restores with varying $h_1$.
}
\end{figure}

\clearpage
\begin{figure}
\centering
\includegraphics[width=11 cm]{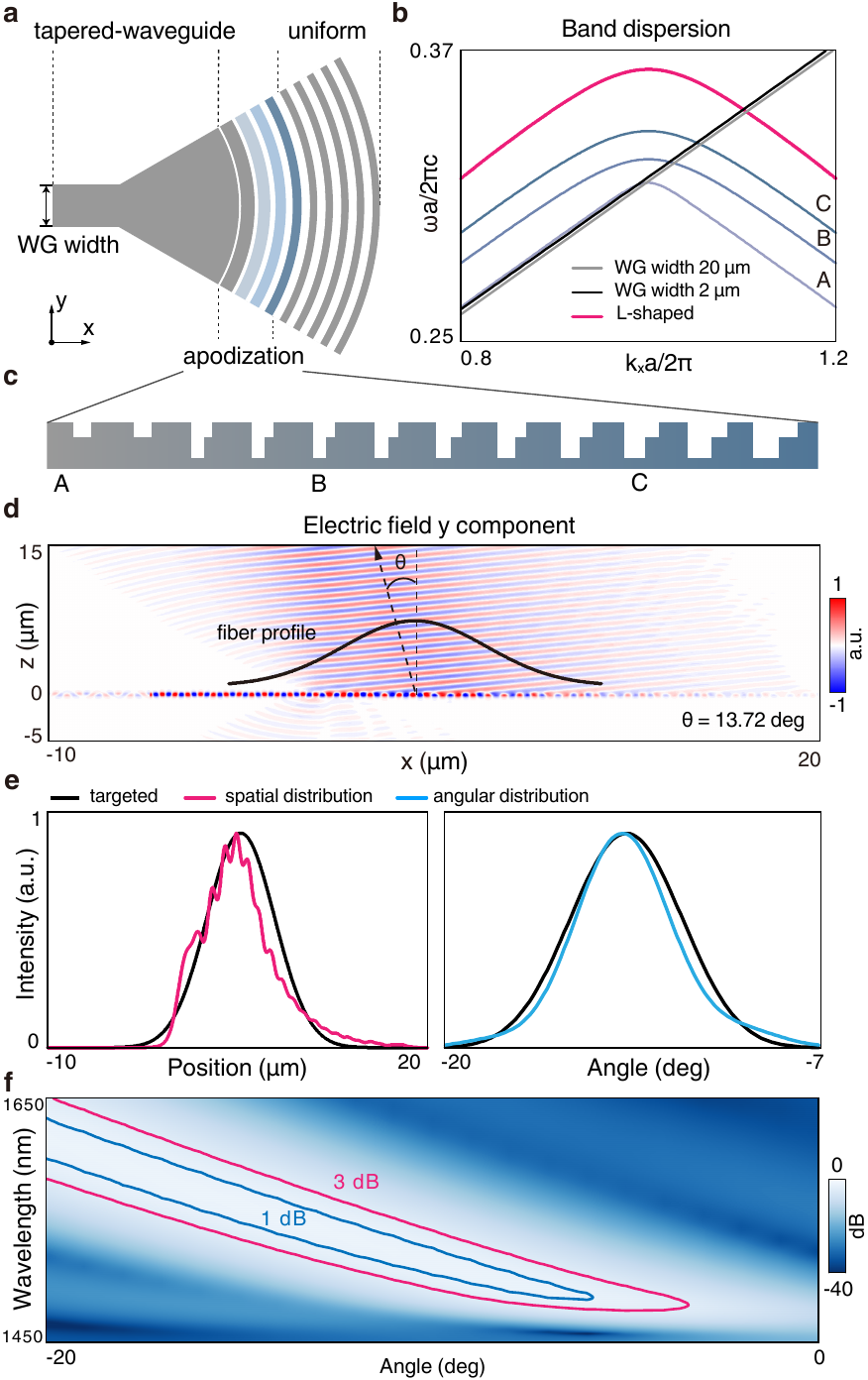}
\end{figure}
\begin{figure}
\caption{\textbf{Design of unidirectional grating coupler}.
\label{mfig2}
(a) 
Top view of the schematic grating coupler which is divided into tapered waveguide, apodization, and uniform regions.
(b) 
Band structures of the grating and waveguides, in which the dispersion of uniform grating adiabatically transits to that of the waveguide. Momentum and group velocity matching is optimized by engineering the apodization region. 
(c) 
The enlarged view of the apodization region.
(d) 
The electrical field distribution ($E_y$) at 1550 nm of the unidirectional grating coupler excited by the mode source. The upward radiation has a $13.72^\circ$ tilt angle that agrees with the UGR.
(e) 
The spatial and angular distributions of the upward radiation demonstrate that the radiating Gaussian beam aligns well with the target fiber.
(f) 
The map of fiber-to-coupler coupling efficiency over the wavelength and alignment angle in which the 1 dB and 3 dB thresholds are marked by blue and red contours. 
}
\end{figure}

\clearpage
\begin{figure}
\centering
\includegraphics[width=17 cm]{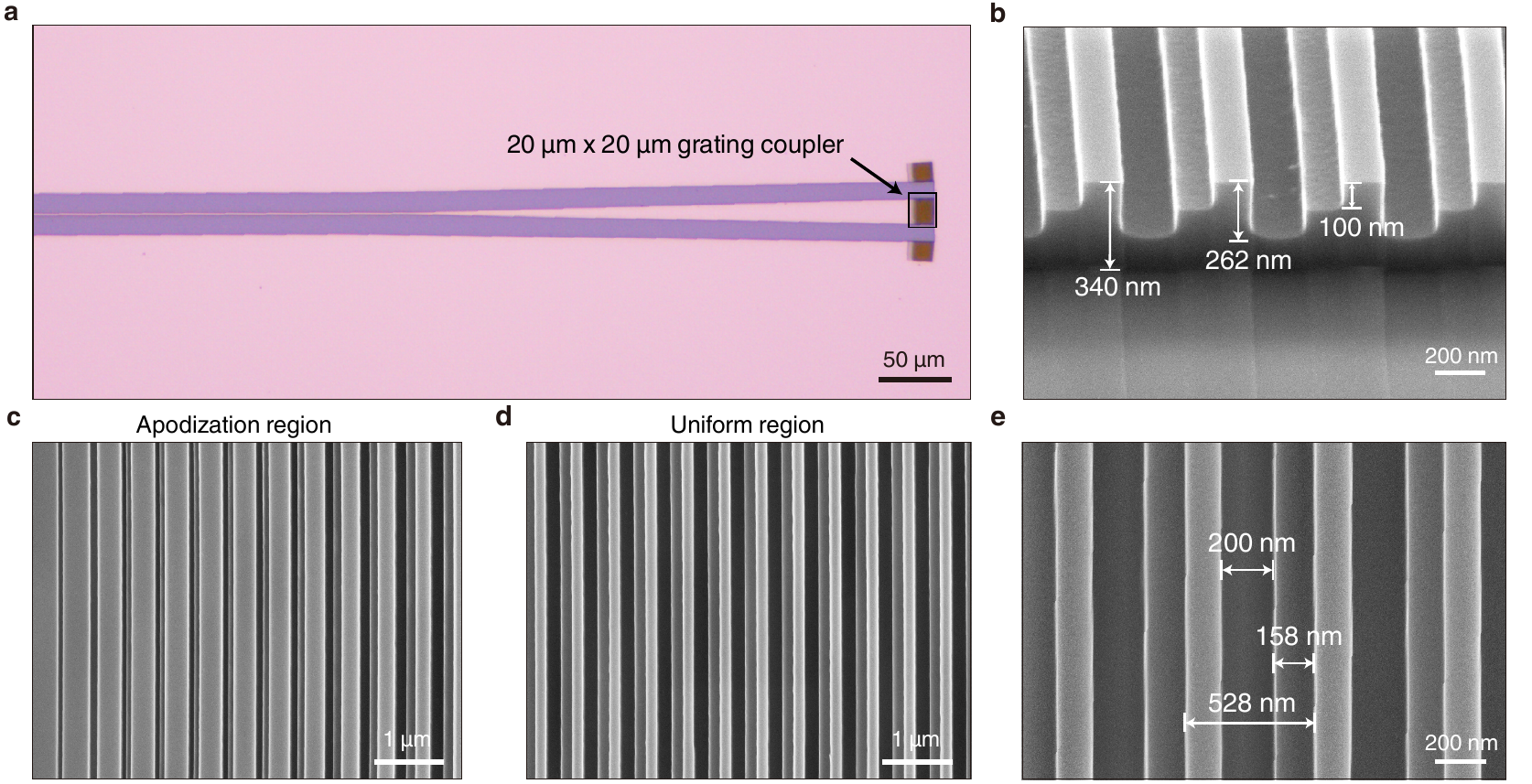}
\caption{\textbf{Sample fabrication}.
\label{mfig3}
(a) 
Optical microscope image of the grating coupler composed of the waveguide, tapered waveguide, and grating coupler with a footprint of 20 $\mu$m$\times$20 $\mu$m.  
(b)
Scanning electron microscope image of the fabricated grating coupler from a side view.
(c,d,e)
Scanning electron microscope images of (c) the apodization region and (d,e) the uniform regions of the grating coupler.
}
\end{figure}

\clearpage
\begin{figure}
\centering
\includegraphics[width=17 cm]{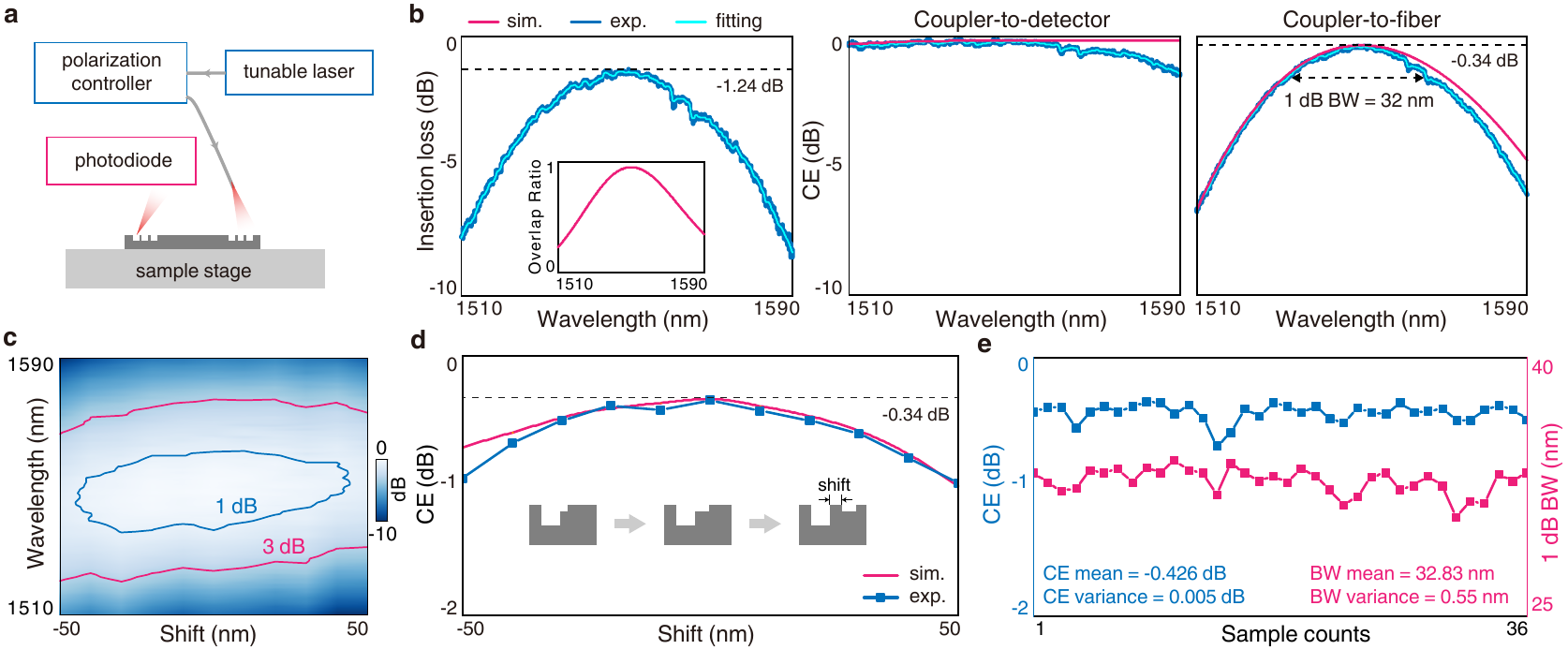}
\caption{\textbf{Characterization of the unidirectional grating coupler }.
\label{mfig4}
(a) 
Schematic of measurement setup for fiber-to-detector optical interconnect.
(b)
Measured insertion loss of the fiber-to-detector link which consists of a pair of unidirectional grating couplers and a waveguide with a length of 7 mm (left). The inset shows the overlap ratio calculated from the mode integral. The spectra of coupler-to-detector (middle) and coupler-to-fiber (right) CEs that are decomposed from the insertion loss and the overlap ratio.
(c)
The map of coupler-to-fiber CEs over wavelength and shift errors with the 1 dB and 3 dB contours labeled as blue and red.
(d)
The peak CEs vs. shift errors.
(e)
The statistics of coupler-to-fiber CEs and 1 dB bandwidth for 36 samples.
}

\end{figure}

\clearpage
\begin{figure}
\centering
\includegraphics[width=12 cm]{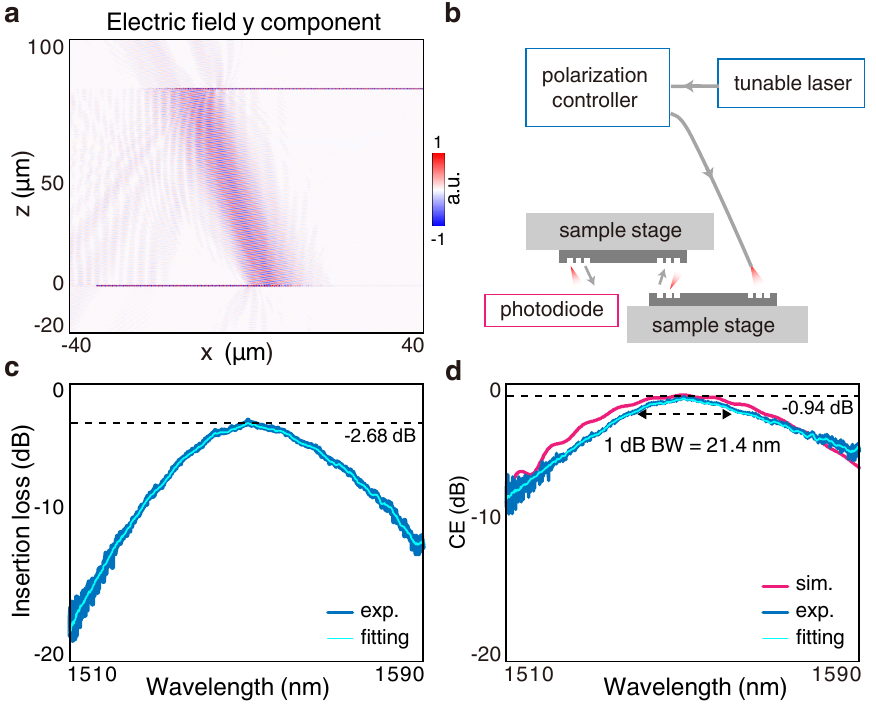}
\caption{\textbf{Characterization of interlay coupling efficiency}.
\label{mfig5}
(a) 
The electrical field distribution at 1550 nm of a pair of interlay grating couplers worked as optical via that vertically connect two photonic chips.
(b)
Schematic of measurement setup for interlay optical interconnect.
(c)
The measured insertion loss of the entire optical link that contains 4 unidirectional grating couplers and a waveguide in a total length of 10 mm, showing a minimal loss of -2.68 dB. 
(d)
The measured interlay coupling efficiency with the peak CE of -0.94 dB and the 1 dB bandwidth of 21.4 nm.
}
\end{figure}

\end{document}